\documentclass[12pt]{article}
\usepackage{amsmath}

\def\nothuge{} 

\begin{document}

\nothuge

\title{\nothuge A Matrix Factorization of Extended Hamiltonian Leads to $N$-Particle Pauli Equation}
\author{\nothuge Irving S. Reed and Todd A. Brun \\ \\
 \nothuge Communication Sciences Institute,\\
 \nothuge University of Southern California,\\
 \nothuge Los Angeles, California 90089-2565}
\date{\nothuge January 10, 2007}
\maketitle

\eject

\begin{abstract}
\nothuge
In this paper the Levy-Leblond procedure for linearizing the Schr\"odinger equation to obtain the Pauli equation for one particle is generalized to obtain an $N$-particle equation with spin.  This is achieved by using the more universal matrix factorization, $G\tilde{G} = |G| I = (-K)^l I$.  Here the square matrix $G$ is linear in the total energy E and all momenta, $\tilde G$ is the matrix adjoint of $G$, $I$ is the identity matrix, $|G|$ is the determinant of $G$, $l$ is a positive integer and $K=H-E$ is Lanczos' extended Hamiltonian where $H$ is the classical Hamiltonian of the electro-mechanical system.  $K$ is identically zero for all such systems, so that matrix $G$ is singular.  As a consequence there always exists a vector function $\underline\theta$ with the property $G\underline\theta=0$.  This factorization to obtain the matrix $G$ and vector function $\underline\theta$ is illustrated first for a one-dimensional particle in a simple potential well.

The quantization of the matrix relation, $G\underline\theta=0$, is performed, using the Planck-de Broglie laws and a Fourier transform to obtain the generalized wave equation, $G(\hat{p},q)\underline\psi(q) = 0$, where $q$ and $\hat p$ are, respectively, the generalized position and momentum-operator vectors of $(n+1)$-dimensions.  This quantization procedure, when applied to the above one-dimensional example, yields immediately a pair of coupled operator equations which become, after eliminating one component of $\underline\psi$, the classical second-order partial-diffferential Schr\"odinger equation of a single one-dimensional particle in a force field.  This same technique, when applied to the classical nonrelativistic Hamiltonian for $N$ interacting particles in an electromagnetic field, is shown to yield for $N=1$ the Pauli wave equation with spin and its generalization to $N$ particles.

Finally this nonrelativistic generalization of the Pauli equation is used to treat the simple Zeeman effect of a hydrogen-like atom as a two-particle problem with spin.  This analysis of the Zeeman effect in a weak magnetic field exhibits the usual two-fold splitting of the energy levels, obtained first by Pauli using the Pauli equation for $N=1$, except for a slight change in the particle mass needed in the Larmor frequency.  This ``increased'' mass, given by $m_L=m_1m_2/(m_2-m_1)$, agrees with the $g$-factor correction for nuclear motion obtained in 1952 by W.H. Lamb using the relativistic, one-particle, Dirac equation with potentials.  Also it is consistent with the known fact that positronium has no weak-field Zeeman effect.
\end{abstract}

\eject

\section{\nothuge Introduction:  Extended Hamiltonian and its Factorization}

The classical Hamiltonian
\begin{equation}
H = H(q_1,\cdots,q_n; p_1,\cdots,p_n; t) ,
\label{eq:1.1}
\end{equation}
which is also called the ``total energy,'' is related to the Lagrangian,
\begin{equation}
L=L(q_1,\cdots,q_n; {\dot q}_1,\cdots,{\dot q}_n; t),
\label{eq:1.2}
\end{equation}
by the relation,
\begin{equation}
H = \sum_{k=1}^n p_k {\dot q}_k - L ,
\label{eq:1.3}
\end{equation}
where the conjugate variables $q_k$ and $p_k$ for ($k=1,2,\cdots,n$) are, respectively, the generalized coordinates and momenta of the particles of a mechanical system in a force field.  The momenta $p_k$ are obtained from the Lagrangian by means of the formulae,
\begin{equation}
p_k = \partial L/\partial {\dot q}_k ,
\label{eq:1.4}
\end{equation}
for $k=1,\ldots,m$ in terms of the time derivatives ${\dot q}_k$ of their {\it conjugate} coordinates $q_k$.  The Hamiltonian $H$ in (\ref{eq:1.1}), as a function of the momenta $p_k$, is obtained from (\ref{eq:1.4}) by solving the equations (\ref{eq:1.4}) for the ${\dot q}_k$, followed by a substitution into (\ref{eq:1.3}), e.g., see [1, Chapt. 6] and [2, Chapt. 8].

In classical mechanics a path minimization of the action integral,
\begin{equation}
A = \int_{t_1}^{t_2} L(q_1,\cdots,q_n; {\dot q}_1,\cdots,{\dot q}_n; t) dt ,
\label{eq:1.5}
\end{equation}
yields Lagrange's equations, a system of $n$ ordinary differential equations for the ``path'' $(q_1(t),q_2(t),\cdots,q_n(t))$ of the system point as a function of time $t$.  However, sometimes it is advantageous for this minimization process to elevate time $t$ to the position of the other coordinates $q_k$.  This means that time is taken to be some monotonically increasing differentiable function,
\begin{equation}
t=t(\alpha),
\label{eq:1.6}
\end{equation}
of some real variable parameter $\alpha$.

In terms of the parameter $\alpha$ in (\ref{eq:1.6}) the action $A$ in (\ref{eq:1.5}) becomes, by an evident change of variables, the integral,
\begin{equation}
A = \int_{t_1}^{t_2} L(q_1,\cdots,q_n; q_1'/t',\cdots,q_n'/t'; t)t' d\alpha ,
\label{eq:1.7}
\end{equation}
where $q_1,\cdots,q_n,t$ are functions of $\alpha$, $q_k'=dq_k/d\alpha$, and $t'=dt/d\alpha$.  In (\ref{eq:1.7}) the original Lagrangian in the integrand of (\ref{eq:1.5}) clearly becomes the new modified Lagrangian,
\begin{eqnarray}
{\bar L} &=& {\bar L}(q_1,\cdots,q_n,q_{n+1}; q_1',\cdots,q_{n+1}') \nonumber\\
&=& L(q_1,\cdots,q_n; q_1'/t',\cdots,q_n'/t'; t)t' ,\nonumber\\
\label{eq:1.8}
\end{eqnarray}
where $q_{n+1}=t$.  Formula (\ref{eq:1.4}) is applied next to ${\bar L}$ in (\ref{eq:1.8}) to obtain the momentum associated with time $t$, namely,
\begin{eqnarray}
p_t &=& \partial{\bar L}/\partial t' \nonumber\\
&=& L + \left( \sum_{k=1}^n \partial L/\partial{\dot q}_k \partial{\dot q}_k/\partial t' \right) t' \nonumber\\
&=& L - \left( \sum_{k=1}^n \partial L/\partial{\dot q}_k q_k'/ {t'}^2 \right) t' \nonumber\\
&=& - \left( \sum_{k=1}^n p_k{\dot q}_k - L \right) .
\label{eq:1.9}
\end{eqnarray}
Hence the momentum $p_t$ is the negative of the total energy defined in (\ref{eq:1.3}), e.g., see [1, Sec. 5.8] and [2, Sec. 8-4].  Note that if the Lagrangian $L$ is not an explicit function of time $t$, then the system is conservative.  In this case $H=E$, where $E$ is a constant.  Thus the momentum $p_t$ in (\ref{eq:1.9}) is given by
\begin{equation}
p_t = - H = - E .
\label{eq:1.10}
\end{equation}
If the system is nonconservative, it is convenient to still let $E$ denote the total energy.  Thus in all cases the relation in (\ref{eq:1.10}), between the momentum $p_t$ and the total energy $E$, must hold.

Now let the new momentum $p_t$ in (\ref{eq:1.9}) be denoted by
\begin{equation}
p_{n+1} = p_t ,
\label{eq:1.11}
\end{equation}
the momentum conjugate to the time coordinate $q_{n+1} = t$.  Also use (\ref{eq:1.3}) to obtain the Hamiltonian,
\begin{equation}
{\bar H} = \sum_{k=1}^{n+1} p_k q'_k - {\bar L} ,
\label {eq:1.12}
\end{equation}
in terms of the modified Lagrangian,
\begin{equation}
{\bar L} = Lq'_{n+1},
\label{eq:1.13}
\end{equation}
the derivatives $q'_1,q'_2,\cdots,q'_{n+1}$ with respect to parameter $\alpha$, and the momenta $p_1,p_2,\cdots,p_{n+1}$, where $p_{n+1}$  is defined in (\ref{eq:1.11}).  A substitution of ${\bar L}$, given in (\ref{eq:1.13}), into (\ref{eq:1.12}), using the Lagrangian $L$ in (\ref{eq:1.3}), yields finally the formula
\begin{eqnarray}
{\bar H} &=& \sum_{k=1}^{n+1} p_k q'_k - \left( \sum_{k=1}^n p_k{\dot q}_k - H \right) q'_{n+1} 
\nonumber\\
&=& (p_t+H)t' \equiv K t',
\label {eq:1.14}
\end{eqnarray}
for the modified Hamiltonian in the parametric system as a function of $p_t$ in (\ref{eq:1.10}) and the original Hamiltonian $H$ in (\ref{eq:1.1}).  By (\ref{eq:1.10}) and the assumption that $t'(\alpha) > 0$ for all $\alpha$ the expression in (\ref{eq:1.14}), namely,
\begin{equation}
K = (p_t+H) = - E + H
\label {eq:1.15}
\end{equation}
is always the constant, zero, for any electro-mechanical system and is called by Lanczos in [1, pg. 190] the extended Hamiltonian function.  Since $K$ does not depend explicitly on the independent variable $\alpha$, every system can be considered to be a conservative system of the extended phase space of $2(n+1)$-dimensions.  Lanczos further points out that any ``fluid particle'' of this phase space is ``steady'' and permanently on the surface,
\begin{equation}
K = {\rm const.} = 0,
\label{eq:1.16}
\end{equation}
of the phase space.  In either a classical or quantum mechanical system the identity in (\ref{eq:1.16}), satisfied by the extended Hamiltonian in ({\ref{eq:1.15}), constitutes the ``fundamental information'' of such systems.

In quantum mechanics the relation in (\ref{eq:1.16}) is used next in a manner similar to the matrix factorization used by Dirac [3, pg. 255] to obtain his famous relativistic wave equation.  To accomplish this we note first by (\ref{eq:1.15}) that the momentum, $p_t=-E$, occurs linearly in $K$ whereas all other momenta $p_1,\cdots,p_n$ generally occur quadratically.  To put the momenta $p_1,\cdots,p_n$ and $p_t$ on the same footing it is desirable, following Dirac [3] and Levy-Leblond [4] later in 1963, in any matrix factorization of $K$ that all momenta occur linearly.

In order to achieve a matrix factorization of $K$ in such a manner that the momenta occur linearly use is made next of the well-known matrix identity,
\begin{equation}
G{\tilde G}=|{\dot G}| I ,
\label{eq:1.17}
\end{equation}
where $|G|$ is the determinant of some square matrix $G$ of complex numbers, ${\tilde G}$ is the matrix adjoint of $G$, sometimes denoted by ${\rm Adj}(G)$, and finally $I$ is the identity matrix.  It is well known that the adjoint of $G$, namely ${\tilde G} = {\rm Adj}(G)$, equals the matrix transpose of the matrix of cofactors of $G$, e.g. see [5, Sec. 1.6].  Thus the universal factorization identity in (\ref{eq:1.17}) is closely related to Cramer's rule for solving a system of linear equations.

Here, using (\ref{eq:1.17}), a matrix $G$ is said to linearly factor the extended Hamiltonian $K$ if the following two conditions hold:

i).  Matrix $G$ is linear in the momenta $p_1,\cdots,p_n$ and $p_t$.

ii).  $|G| = (-K)^l$ where $l$ is a positive integer.

To illustrate the linear factorization in i) and ii) an example of the extended Hamiltonian for a one-dimensional particle in a force-field is given next.
\\\\
{\bf Example 1}  Let a one-dimensional particle of mass $m$ be in a time-varying force field of potential $V(x,t)$.  Then the Hamiltonian is
\begin{equation}
H(x,t;p) = p^2/2m + V(x,t) ,
\label{eq:1.18}
\end{equation}
where $p$ is the momentum of the particle.  Thus by (\ref{eq:1.15}) and (\ref{eq:1.16}) the negative of the extended Hamiltonian for this system is given by
\begin{eqnarray}
-K(x,t;p,-E) &=& E - H(x,t;p) \nonumber\\
&=& E - V - p^2/2m \nonumber\\
&=& 0 .
\label{eq:1.19}
\end{eqnarray}

If one lets $G$ in (\ref{eq:1.17}) be the $2\times2$ matrix,
\begin{equation}
G = \left[ \begin{array}{cc}
E-V & p/\sqrt{2m} \\
p/\sqrt{2m} & 1
\end{array} \right] ,
\label{eq:1.20}
\end{equation}
then its determinant is computed easily to be
\begin{equation}
|G| = -K(x,t;p,-E) = 0
\label{eq:1.21}
\end{equation}
in terms of the extended Hamiltonian in (\ref{eq:1.19}).  Evidently by (\ref{eq:1.20}) and (\ref{eq:1.21}) the above conditions i) and ii) are satisfied.  Hence matrix $G$ in (\ref{eq:1.20}) is an example of a liner factorization of $-K$ in (\ref{eq:1.19}).  Other such factorizations of $K$ include the non-symmetric matrices,
\[
G = \left[ \begin{array}{cc}
E-V & bp \\
ap & 1
\end{array} \right] ,
\]
for $a\cdot b=1/2m$.  However, it appears that $K$ can be factored usually into Hermitian matrices.  Thus for simplicity, though not necessary, it is assumed henceforth that any matrix factorization $G$ of $K$ is Hermitian as well as linear in the momenta.

The fact that the determinant $|G|$ of $G$ in (\ref{eq:1.20}) is zero implies that $G$ is a {\it singular} Hermitian matrix.  But it is also well known in general that any matrix $G$ is singular if and only if there exists some nonzero vector $\underline\theta$ such that
\begin{equation}
G\underline\theta = 0 .
\label{eq:1.22}
\end{equation}
Thus, in particular, the relation $K=0$, in (\ref{eq:1.19}) holds if and only if the relation,
\begin{equation}
G\underline\theta = \left[ \begin{array}{cc}
E-V & p/\sqrt{2m} \\
p/\sqrt{2m} & 1
\end{array} \right]
\left[ \begin{array}{c}
\theta_1 \\
\theta_2
\end{array} \right]  = 0 ,
\label{eq:1.23}
\end{equation}
or that the two simultaneous equations,
\begin{eqnarray}
(E-V)\theta_1 + \left(p/\sqrt{2m}\right)\theta_2 &=& 0, \nonumber\\
\left(p/\sqrt{2m}\right)\theta_1 +  \theta_2 &=& 0 ,
\label{eq:1.24}
\end{eqnarray}
hold for some vector $\underline\theta$.

Since $K=0$, the most general solution of the pair of equations in (\ref{eq:1.24}) is given by
\begin{eqnarray}
\theta_1 &=& \lambda , \nonumber\\
\theta_2 &=& -\left(p/\sqrt{2m}\right)\lambda ,
\label{eq:1.25}
\end{eqnarray}
where $\lambda$ is an arbitrary function of $x,t,p,E$.  Thus all vectors,
\begin{equation}
\underline\theta =\left[ \begin{array}{c}
 \lambda \\
 -\left(p/\sqrt{2m}\right)\lambda 
\end{array} \right] ,
\label{eq:1.26}
\end{equation}
such that $\lambda$ is an arbitrary function, satisfy the null solution in (\ref{eq:1.22}).

It is demonstrated in the next section that the Fourier transform of $\underline\theta$ in (\ref{eq:1.25}) with respect to $p$ is, in fact, closely related to the $\psi$-function of the standard Schr\"odinger equation of a particle in a potential well.  Thus the fact, that the factorization process of $K$ always yields a singular matrix $G$ and a nonzero solution $\underline\theta$ of $G\underline\theta=0$, helps to supply an underlying basis for the Schr\"odinger $\psi$-function of quantum mechanics.

\section{\nothuge Fourier-Momentum Transform of $G{\underline\theta} = 0$ Using The Planck-de Broglie Laws}

In this section the Planck and de~Broglie laws,
\begin{equation}
E = \hbar\omega \ \ {\rm and}\ \ {\underline p} = \hbar {\underline k} ,
\label{eq:2.1}
\end{equation}
imply that the relation, $G{\underline\theta}=0$, in (\ref{eq:1.22}) belongs to the domain of frequencies, $\omega$ and $\underline k$.  This suggests that the energy and momenta in (\ref{eq:1.22}) be Fourier transformed to the spatial domain.  Here in (\ref{eq:2.1}) $\hbar = h/2\pi$, where $h$ is Planck's constant, $\omega$ is angular frequency in radians per second, ${\underline p}$ is the momentum vector of some particle and $\underline k$ is the wave-number vector in radians per unit distance associated with the particle.  To perform such a Fourier transform it is convenient to assume that the generalized coordinates in (\ref{eq:1.3}) are Cartesian coordinates, e.g., $q_1, q_2, \cdots, q_n$ are the components of the position vectors of the $N$ particles, so that $n=3N$.

Now note that the set $N_G$ of all solutions ${\underline\theta}$ of the relation, $G{\underline\theta}=0$, in (\ref{eq:1.22}) is called the {\it null space} of the Hermitian matrix $G$.  Also by (\ref{eq:2.1}) any solutions ${\underline\theta}$ in $N_G$ can be expressed in the abbreviated form,
\begin{equation}
{\underline\theta} = {\underline\theta}({\underline p}, {\underline q} ) = {\underline\theta}(\hbar{\underline k}, {\underline q} ) ,
\label{eq:2.2}
\end{equation}
where
\begin{equation}
{\underline q} = (q_1, \cdots, q_n, q_{n+1})
\label{eq:2.3}
\end{equation}
and
\begin{eqnarray}
{\underline p} &=& (p_1, \cdots, p_n, p_{n+1}) , \nonumber\\
{\underline k} &=& (k_1, \cdots, k_n, k_{n+1})
\label{eq:2.4}
\end{eqnarray}
are the generalized position, momentum and wave-number vectors with the $(n+1)$th coordinates given, respectively, by
\begin{eqnarray}
q_{n+1} &=& t, \nonumber\\
p_{n+1} &=& - E, \ \ {\rm and} \nonumber\\
k_{n+1} &=& - \omega .
\label{eq:2.5}
\end{eqnarray}

Since the matrix,
\begin{equation}
G = G(\underline{p},\underline{q}),
\label{eq:2.6}
\end{equation}
in (\ref{eq:1.17}) is assumed to be the result of a Hermitian matrix factorization of $K$, all solutions $\underline\theta$ of (\ref{eq:1.22}) automatically satisfy the constraint, $K=0$, in (\ref{eq:1.10}).  Hence vectors $\underline{p}$ and $\underline{q}$ in the function $\underline\theta$ are independent vectors, e.g. see [1, Secs. 6.9 and 6.10].  This latter fact makes it possible to consider $\underline{q}$ to be a parameter vector of the system.

Classically the equations of motion for vectors $\underline{p}$ and $\underline{q}$ are determined completely from the Hamiltonian by the Hamilton-Jacobi equations.  However, particles of a system, which satisfy the quantum rules in (\ref{eq:2.1}), are wave-like in nature.  Hence, the components of the generalized wave-number vector $\underline{k}$ in (\ref{eq:2.4}) are the appropriate observables of the system.  By (\ref{eq:2.2}) any solution $\underline\theta$ of (\ref{eq:1.22}) is a function of $\underline{k}$ and $\underline{q}$ as well as a function of $\underline{p}$ and $\underline{q}$.  Thus $\underline\theta(\underline{p},\underline{q})$ can be considered to be some wave-number and frequency spectrum of the electro-mechanical system with $\underline{q}$ being a parameter vector.

Explicitly this wave function is a $(n+1)$-dimensional Fourier transform of $\underline\theta(\underline{p},\underline{q})$ of the form,
\begin{equation}
\underline\phi(\underline{q},\underline{q'}) = \int_{R_{n+1}} \underline\theta(\underline{p},\underline{q}) e^{\frac{i}{\hbar}\underline{p}\cdot\underline{q'}} dV_{\underline{p}} ,
\label{eq:2.7}
\end{equation}
where $dV_{\underline{p}}$ is the volume element of the $(n+1)$-dimensional Cartesian space $R_{n+1}$.  Here also $\underline{q'}$ is the position of the system, associated with this Fourier transform of function $\underline\theta(\underline{p},\underline{q})$, and $\underline{q}$ is the position vector in $\underline\theta(\underline{p},\underline{q})$ that is left unchanged by the mapping.

Now apply the same Fourier transform in (\ref{eq:2.7}) to the relation in (\ref{eq:1.22}) to obtain
\begin{equation}
 \int_{R_{n+1}} e^{\frac{i}{\hbar}\underline{p}\cdot\underline{q'}} G\underline\theta dV_{\underline{p}} = 0.
\label{eq:2.8}
\end{equation}
Also remember that matrix $G$ was chosen in the last section to be linear in the components of the momentum vector $\underline{p}$.  Hence to evaluate the above relation one needs integrals only of the form,
\begin{equation}
 \int_{R_{n+1}} p_k e^{\frac{i}{\hbar}\underline{p}\cdot\underline{q'}} \underline\theta(\underline{p},\underline{q}) dV_{\underline{p}} = \frac{i}{\hbar} \frac{\partial}{\partial q_k'} \underline\phi(\underline{q},\underline{q'}) ,
\label{eq:2.9}
\end{equation}
for $k=1,\cdots,n+1$ where $\underline\phi(\underline{q},\underline{q'})$ is defined in (\ref{eq:2.7}).  Finally a use of (\ref{eq:2.9}) and (\ref{eq:2.6}) transforms (\ref{eq:2.8}) into
\begin{equation}
G(\underline{\hat{p}'},\underline{q}) \underline\phi(\underline{q},\underline{q'}) = 0 ,
\label{eq:2.10}
\end{equation}
where $\underline{\hat{p}'}$ is the generalized momentum-operator vector,
\[
\underline{\hat{p}'} = \frac{\hbar}{i} \left( \frac{\partial}{\partial q_1'}, \cdots , \frac{\partial}{\partial q_{n+1}'} \right) .
\]

To obtain a general Schr\"odinger-like wave equation which is a function of only a single position vector, one sets $\underline{q'} = \underline{q}$ in (\ref{eq:2.10}).  This yields
\begin{equation}
G(\underline{\hat{p}'},\underline{q}) \underline\phi(\underline{q},\underline{q'}) \Biggr|_{\underline{q'}=\underline{q}} = 0 ,
\label{eq:2.11}
\end{equation}
or the vector wave equation,
\begin{equation}
G(\underline{\hat{p}},\underline{q}) \underline\psi(\underline{q})  = 0 ,
\label{eq:2.12}
\end{equation}
of the position vector $\underline{q}$ and the quantum-operator vector,
\begin{equation}
\underline{\hat{p}} = \frac{\hbar}{i} \left( \frac{\partial}{\partial q_1} , \cdots , \frac{\partial}{\partial q_{n+1}} \right) ,
\label{eq:2.13}
\end{equation}
where
\begin{equation}
\underline\psi(\underline{q})  = \underline\phi(\underline{q},\underline{q}) ,
\label{eq:2.14}
\end{equation}
is a vector, Schr\"odinger amplitude function, of the $n+1$ components of $\underline{q}$.  The general formula in (\ref{eq:2.12}) for a wave equation is applied next to the matrix factor $G$ defined in (\ref{eq:1.20}) of Example 1.
\\\\
{\bf Example 2}  The use of (\ref{eq:1.20}) in (\ref{eq:2.12}) yields the operator-matrix identity,
\begin{equation}
\left[ \begin{array}{cc} \hat{E} - V & \hat{p}/\sqrt{2m} \\
\hat{p}/\sqrt{2m} & 1 \end{array} \right] 
\left[ \begin{array}{c} \psi_1 \\ \psi_2 \end{array} \right] = 0
\label{eq:2.15}
\end{equation}
or the pair of coupled operator equations,
\begin{eqnarray}
(\hat{E} - V) \psi_1 &+& (\hat{p}/\sqrt{2m}) \psi_2 = 0 \nonumber\\
(\hat{p}/\sqrt{2m}) \psi_1 &+& \psi_2 = 0 , 
\label{eq:2.16}
\end{eqnarray}
where
\begin{equation}
\hat{p} = \frac{\hbar}{i}\frac{\partial}{\partial x} \ \ {\rm and} \ \ \hat{p}_t = - \hat{E} = \frac{\hbar}{i}\frac{\partial}{\partial t} .
\label{eq:2.17}
\end{equation}
Solving the second equation in (\ref{eq:2.16}) for $\psi_2$ and substituting into the first equation yields immediately
\begin{equation}
(\hat{p}/2m + V)\psi_1 = \hat{E}\psi_1
\label{eq:2.18}
\end{equation}
or by (\ref{eq:2.17}) the wave equation
\begin{equation}
\left( - \hbar^2 \frac{\partial^2}{\partial x^2} + V \right) \psi_1 = i\hbar\frac{\partial\psi_1}{\partial t} ,
\label{eq:2.19}
\end{equation}
for the Schr\"odinger partial-differential equation of second order of a single one-dimensional particle in a force field in standard form.

The above procedures, used in Examples 1 and 2, to factor the extended Hamiltonian and to find its associated Schr\"odinger equation, are applied in the next section to factor the extended Hamiltonian of an $N$-particle system in an electromagnetic field.  Formula (\ref{eq:2.12}) is utilized then to find a generalization of the Pauli equation from one particle to $N$ particles.

\section{\nothuge Factorization of the Non-Relativistic Extended Hamiltonian for $N$ particles}

The relativistic Hamiltonian for $N$ charged particles with magnetic potentials is known, e.g., see [6, Lec. 2], [7, Chapt. 12] and [2, Chapts. 7 and 8].  It is given by
\begin{eqnarray}
H &=& \sum_{k=1}^N \left( \sqrt{m_k^2c^4 + (c{\underline p}_k - q_k{\underline A}_k)^2} - m_k c^2 \right)\nonumber\\
&& \ \ \ \ \  +\  U\nonumber\\
&=& E ,
\label{eq:3.1}
\end{eqnarray}
where $U$ is the scalar potential energy of the $N$ particles due to external and interacting electric fields as well as other forces acting on the system, and $q_k$ is the cahrge of the $k$-th particle.  Here as well ${\underline A}_k$ is the vector potential of the magnetic field impinging on the $k$-th particle, including externally imposed and mutually interacting magnetic fields of the $N$ charged particled due to motion.

The non-relativistic Hamiltonian of the system can be found directly by classical methods or by approximating the relativistic Hamiltonian as follows:
\begin{equation}
\begin{split}
H =& 
\sum_{k=1}^N \Biggl( m_kc^2  \Biggl[  1 + \frac{1}{m_k^2c^4}(c{\underline p}_k - q_k{\underline A}_k)^2 + \cdots \Biggr] 
 - m_kc^2 \Biggr) + U \\
=&  \sum_{k=1}^N \frac{1}{2m_k}  \left( {\underline p}_k - \frac{q_k}{c}{\underline A}_k \right)^2 + U . \nonumber
\end{split}
\end{equation}
Subtracting $H$ from the total energy $E$ yields the desired nonrelativistic extended Hamiltonian nullity relation,
\begin{equation}
-K \equiv E - U - \sum_{k=1}^N \frac{1}{2m_k} \left( {\underline p}_k - \frac{q_k}{c}{\underline A}_k \right)^2 = 0 ,
\label{eq:3.2}
\end{equation}
for the $N$-particle system, where $K$ is defined in (\ref{eq:1.15}).  The higher order terms in the above expansion of (\ref{eq:3.1}) can be used to find relativistic corrections to (\ref{eq:3.2}).

It is demonstrated in the ensuing discussion that the determinant of the following $2^N(N+1)\times2^N(N+1)$ Hermitian symmetric matrix,
\begin{equation}
G = \left[
\begin{array}{cccc}
(E-U)I & \frac{P_1 - \frac{q_1A_1}{c}}{\sqrt{2m_1}}  & \cdots  & \frac{P_N - \frac{q_NA_N}{c}}{\sqrt{2m_N}}  \\
\frac{P_1 - \frac{q_1A_1}{c}}{\sqrt{2m_1}} & I  & \cdots  & 0  \\
\vdots & \vdots  & \ddots  & \vdots  \\
\frac{P_N - \frac{q_NA_N}{c}}{\sqrt{2m_N}} & 0  & \cdots  & I  
\end{array}
\right]
\label{eq:3.3}
\end{equation}
is given by
\begin{equation}
\det(G) = K^2 ,
\label{eq:3.4}
\end{equation}
where $K$ is the extended Hamiltonian relation, and $I$ and $0$ represent the unit and zero $2^N\times2^N$ matrices.  In (\ref{eq:3.3}) $P_j$ and $A_j$ for $j=1,2,\cdots,N$ are defined respectively by the Hermitian matrices,
\begin{equation}
P_j = \hat{\underline\sigma}_j \cdot{\underline p_j} \ \ {\rm and}\ \ A_j = \hat{\underline\sigma}_j \cdot{\underline A_j} ,
\label{eq:3.5}
\end{equation}
with
\begin{equation}
\hat{\underline\sigma}_j = [\sigma_{1j},\sigma_{2j},\sigma_{3j}]
\label{eq:3.6}
\end{equation}
being the spin-vector operators for $N$ independent particles.  These operators are defined by the commutation relations
\begin{equation}
\begin{split}
[\sigma_{1j},\sigma_{2k}] = i \delta_{jk} \sigma_{3j} , \\
[\sigma_{2j},\sigma_{3k}] = i \delta_{jk} \sigma_{1j} , \\
[\sigma_{3j},\sigma_{1k}] = i \delta_{jk} \sigma_{2j} .
\end{split}
\end{equation}
with $\delta_{jk}$ the Kronecker delta.  They are traceless and idempotent:
\begin{equation}
\sigma_{1j}^2 = \sigma_{2j}^2 = \sigma_{3j}^2 = I .
\end{equation}
For a single spin, these operators can be represented by the standard $2\times2$ Pauli spin matrices:
\begin{eqnarray}
\sigma_1 &=& \left[
\begin{array}{cc}
0  & 1  \\
1  & 0  
\end{array}
\right] , \nonumber\\
\sigma_2 &=& \left[
\begin{array}{cc}
0  & -i  \\
i  & 0  
\end{array}
\right] , \nonumber\\
\sigma_3 &=& \left[
\begin{array}{cc}
1  & 0  \\
0  & -1  
\end{array}
\right] .
\label{eq:3.7}
\end{eqnarray}
By (\ref{eq:3.5}), (\ref{eq:3.6}) and (\ref{eq:3.7}) one would obtain more explicitly
\begin{eqnarray}
P &=& \left[
\begin{array}{cc}
p_3  & p_1-ip_2  \\
p_1+ip_2  & -p_3  
\end{array}
\right] \ \ {\rm and} \nonumber\\
A &=& \left[
\begin{array}{cc}
A_3  & A_1-iA_2  \\
A_1+iA_2  & -A_3  
\end{array}
\right] ,
\label{eq:3.8}
\end{eqnarray}
for a one-particle system.  For $N$ particles, the spin operators become $2^N\times2^N$ matrices, which we can represent by
\begin{eqnarray}
\sigma_{1j} &=& \underbrace{I_2 \otimes \cdots I_2}_{j-1 \ {\rm times}} \otimes \sigma_1 \otimes
\underbrace{I_2 \otimes \cdots I_2}_{N-j \ {\rm times}}  \nonumber\\
&\equiv& (I_2)^{\otimes j-1} \otimes \sigma_1 \otimes (I_2)^{\otimes N-j} \nonumber\\
&=& I_{2^{j-1}} \otimes \sigma_1 \otimes I_{2^{N-j}} ,
\label{tensor_product_spin}
\end{eqnarray}
and similarly for $\sigma_{2j}$ and $\sigma_{3j}$, where $\otimes$ denotes the tensor (or Kronecker) product, $I_n$ is the $n\times n$ identity matrix, and $\sigma_{1,2,3}$ are the $2\times2$ Pauli matrices from (\ref{eq:3.7}).  By inserting these definitions into (\ref{eq:3.5}), we get $2^N\times2^N$ matrices for $P_j$ and $A_j$, which give us the $2^N\times2^N$ sub-block matrices in matrix $G$ in (\ref{eq:3.3}) for $j=1,2,\cdots,N$.

To find the determinant $|G|$ of $G$ in (\ref{eq:3.4}) first simplify the notation of matrix $G$ in (\ref{eq:3.3}) by defining the following $2^N\times2^N$ sub-matrices of $G$, namely,
\begin{eqnarray}
aI &\equiv& (E-U)I \ \ {\rm and} \nonumber\\
H_j &=& \frac{1}{\sqrt{2m_j}} \left( P_j - \frac{q_j}{c} A_j \right) ,
\label{eq:3.9}
\end{eqnarray}
for $j=1,2,\cdots,N$, where $a$ is the scalar function $E-U$.  Then matrix $G$ is re-expressible in the simpler sub-block form as follows:
\begin{eqnarray}
G &=& \left[
\begin{array}{cccc}
aI & H_1  & \cdots  & H_N  \\
H_1 & I  & \cdots  & 0  \\
\vdots & \vdots  & \ddots  & \vdots  \\
H_N &  0 &  \cdots & I  
\end{array}
\right] \nonumber\\
&=& \left[
\begin{array}{cc}
aI_{2^N}  & B  \\
B^\dagger  & I_{N2^N}  
\end{array}
\right] ,
\label{eq:3.10}
\end{eqnarray}
where $I_{2^N}$ and $I_{N2^N}$ are $2^N\times2^N$ and $(N2^N)\times(N2^N)$ unit matrices.  Also in (\ref{eq:3.10}) $B$ and $B^\dagger$ are, respectively, the $2^N\times(N2^N)$ rectangular matrix,
\begin{equation}
B = [H_1, H_2, \cdots, H_N]
\label{eq:3.11}
\end{equation}
and its conjugate transpose.

Note now that matrix $G$ in (\ref{eq:3.10}) has the form of the general $n\times n$ matrix,
\begin{equation}
M = \left[
\begin{array}{cc}
A  & B  \\
C  & D  
\end{array}
\right] ,
\label{eq:3.12}
\end{equation}
where $A$,$B$,$C$ and $D$ are, respectively, $k\times k$, $k\times(n-k)$, $(n-k)\times k$, and $(n-k)\times(n-k)$ matrices.  It is shown in Appendix A that if the sub-block matrix $D$ in (\ref{eq:3.12}) is nonsingular, then
\begin{equation}
|M| = |D| \cdot |A - B D^{-1} C|
\label{eq:3.13}
\end{equation}
is the determinant of the $n\times n$ matrix $M$.  Thus is (\ref{eq:3.12}) if one lets $n=2^N(N+1)$, $k=2^N$ and makes the correspondences,
\[
A \leftrightarrow aI_{2^N} , \ \ 
B \leftrightarrow B ,
\]
\[
C \leftrightarrow B^\dagger , \ \ 
D \leftrightarrow I_{N2^N} ,
\]
then by (\ref{eq:3.13}) one obtains
\begin{equation}
|G| = |aI_{2^N} - B B^\dagger| .
\label{eq:3.14}
\end{equation}
But by (\ref{eq:3.11}) and (\ref{eq:3.9}) one has
\begin{equation}
BB^\dagger = \sum_{j=1}^N H_j^2 .
\label{eq:3.15}
\end{equation}
Also by (\ref{eq:3.9}) the $H_j^2$ in (\ref{eq:3.15}) are evaluated by the matrix squarings,
\begin{eqnarray}
H_j^2 &=& \left[ h_{1j}\sigma_{1j}+h_{2j}\sigma_{2j} + h_{3j}\sigma_{3j}\right]^2 \nonumber\\
&=& h_j^2 I_{2^N} ,
\label{eq:3.16}
\end{eqnarray}
where
\begin{eqnarray}
h_j^2 &=& \frac{1}{2m_j} \left( {\underline p}_j - \frac{q_j}{c}{\underline A}_j \right) \cdot \left( {\underline p}_j - \frac{q_j}{c}{\underline A}_j \right) \nonumber\\
&=& \frac{1}{2m_j} \left( {\underline p}_j - \frac{q_j}{c}{\underline A}_j \right)^2
\label{eq:3.17}
\end{eqnarray}
is the square of the magnitude of the vector $\frac{1}{\sqrt{2m_j}} \left( {\underline p}_j - \frac{q_j}{c}{\underline A}_j \right)$, for $j=1,2,\cdots,N$.  A substitution of (\ref{eq:3.16}) into (\ref{eq:3.15}) yields finally the desired determinant of $G$ in (\ref{eq:3.8}), namely,
\begin{eqnarray}
|G| &=& \left| \left( a - \sum_{j=1}^N h_j^2 \right) I \right| \nonumber\\
&=& \left( a - \sum_{j=1}^N h_j^2 \right)^2 \nonumber\\
&=& \left[ E - U - \sum_{j=1}^N \frac{1}{2m_j} \left( {\underline p}_j - \frac{q_j}{c}{\underline A}_j \right)^2 \right]^2 \nonumber\\
&=& K^2 = 0 .
\label{eq:3.18}
\end{eqnarray}

Evidently the result in (\ref{eq:3.18}) for the Hermitian matrix $G$ in (\ref{eq:3.3}) satisfies conditions i) and ii) in Sec. 1 for $G$ to linearly factor the extended Hamiltonian $K$ in (\ref{eq:3.2}).  Thus by the arguments in Sec. 2 the quantum wave equation, associated with $K$ in (\ref{eq:3.2}) and its matrix factor $G$ in (\ref{eq:3.3}), can be obtained by a use of the formula in (\ref{eq:2.11}).  To accomplish this note first by (\ref{eq:3.18}) that since the matrix $G$ is singular, there exists a nonzero solution vector 
${\underline\theta}({\underline p}_1,\cdots,{\underline p}_N,-E; {\underline q}_1,\cdots,{\underline q}_N, t)$ of the equation
\begin{equation}
G{\underline\theta} = 0,
\label{eq:3.19}
\end{equation}
for the momentum vectors, ${\underline p}=({\underline p}_1,{\underline p}_2,\cdots,{\underline p}_N,-E)$ in (\ref{eq:2.4}) and the position vectors ${\underline q} = ({\underline q}_1,\cdots,{\underline q}_N,t)$ in (\ref{eq:2.5}).

Next for the vector ${\underline\theta}$ in (\ref{eq:3.19}) let
\begin{equation}
{\underline\theta} = \left[
\begin{array}{c}
{\underline\theta}_1   \\
\vdots   \\
{\underline\theta}_{N+1}
\end{array}
\right]
\label{eq:3.20}
\end{equation}
be the vector of the $(N+1)$ $2^N$-component vectors, 
for $k=1,2,\cdots,N+1$.  The vectors ${\underline\theta}_k$  in (\ref{eq:3.20})
are {\it generalized spinor vectors} or more simply, spinors.  A substitution of the vectors in (\ref{eq:3.20}) of spinors into (\ref{eq:3.19}), where $G$ is given in (\ref{eq:3.3}), yields the following system of linear equations,
\begin{eqnarray}
(E-U){\underline\theta}_1 + \frac{1}{\sqrt{2m_1}} \left(P_1 - \frac{q_1}{c} A_1\right) {\underline\theta}_2 && \nonumber\\
+ \cdots + \frac{1}{\sqrt{2m_N}} \left(P_N - \frac{q_N}{c} A_N\right) {\underline\theta}_{N+1} &=& 0 \nonumber\\
\frac{1}{\sqrt{2m_1}} \left(P_1 - \frac{q_1}{c} A_1\right) {\underline\theta}_1 + {\underline\theta}_2 &=& 0 \nonumber\\
\vdots \ \ \ \ \ \ && \nonumber\\
\frac{1}{\sqrt{2m_N}} \left(P_N - \frac{q_N}{c} A_N\right) {\underline\theta}_1 + {\underline\theta}_{N+1} &=& 0 ,\nonumber\\
\label{eq:3.22}
\end{eqnarray}
in the $N+1$ spinors ${\underline\theta}_1, {\underline\theta}_2, \cdots, {\underline\theta}_{N+1}$, defined in (\ref{eq:3.20}).  Thus in terms of these $N+1$ spinor vectors the extended Hamiltonian relations, $K=0$, in (\ref{eq:1.16}) is valid if and only if the linear system in (\ref{eq:3.22}) has non-zero solutions for spinors ${\underline\theta}_1, {\underline\theta}_2, \cdots, {\underline\theta}_{N+1}$ for all momenta ${\underline p}_1,\cdots,{\underline p}_N$ and total energy $E$.  

The null relation in (\ref{eq:3.19}) for matrix $G$ in (\ref{eq:3.3}) transforms immediately by the procedures of the last section into the wave equation,
\begin{equation}
G\left({\hat{\underline p}},{\underline q} \right) {\underline\psi}({\underline q}) = 0,
\label{eq:3.23}
\end{equation}
in (\ref{eq:2.12}), where
\begin{equation}
{\hat{\underline p}} = ({\hat{\underline p}}_1, \cdots, {\hat{\underline p}}_N,-{\hat E})
\label{eq:3.24}
\end{equation}
is the generalized momentum-operator vector with
\begin{equation}
{\hat{\underline p}}_k = \frac\hbar{i}\nabla_k \ \ {\rm and}\ \ {\hat E} = i\hbar \frac{\partial}{\partial t}
\label{eq:3.25}
\end{equation}
for $k=1,2,\cdots,N$ being three-dimensional momentum operator vectors and the energy operator of quantum mechanics.  Finally a use of (\ref{eq:3.25}) in (\ref{eq:3.5}) and then a substitution into (\ref{eq:3.23}) yields the ``transformed'' linear system,
\begin{eqnarray}
\left( \hat{E} - U\right) {\underline\psi}_1
  + \frac{1}{\sqrt{2m_1}} \left( \hat{P}_1 - \frac{q_1}{c} A_1 \right) {\underline\psi}_2 && \nonumber\\
  + \cdots
  + \frac{1}{\sqrt{2m_N}} \left( \hat{P}_N - \frac{q_N}{c} A_N \right) {\underline\psi}_{N+1} &=& 0 \nonumber\\
\frac{1}{\sqrt{2m_1}} \left( \hat{P}_1 - \frac{q_1}{c} A_1 \right) {\underline\psi}_1
  + {\underline\psi}_2 &=& 0 \nonumber\\
\vdots \ \ \ \ \ \ && \nonumber\\
\frac{1}{\sqrt{2m_N}} \left( \hat{P}_N - \frac{q_N}{c} A_N \right) {\underline\psi}_1
  + {\underline\psi}_{N+1} &=& 0 ,\nonumber\\
\label{eq:3.26}
\end{eqnarray}
of $(N+1)$ coupled partial-differential equations for the $(N+1)$ spinors, defined in (\ref{eq:3.20}),
where
\begin{eqnarray}
\hat{P}_k &=& {\underline\sigma}\cdot\hat{\underline{p}}_k = \frac{\hbar}{i} {\underline\sigma} \cdot \nabla_k \ \ {\rm and} \nonumber\\
\hat{E} &=& i\hbar\frac{\partial}{\partial t}
\label{eq:3.27}
\end{eqnarray}
for $k=1, 2, \cdots, N$, are momentum-operator matrices and the energy operator of quantum mechanics.

To convert the system in (\ref{eq:3.27}) of linear, coupled, partial-differential equations for the $N+1$ spinors ${\underline\psi}_1, \cdots, {\underline\psi}_{N+1}$ into a single second-order, Schr\"odinger-like, equation in the single spinor ${\underline\psi}_1$ first solve the $k$-th equation of the last $N$ equations for ${\underline\psi}_k$ to obtain
\begin{equation}
{\underline\psi}_k = - \frac{1}{\sqrt{2m_k}} \left( \hat{P}_k - \frac{q_k}{c} A_k \right) {\underline\psi}_1
\label{eq:3.28}
\end{equation}
for $k=2, 3, \cdots, N+1$.  Finally a substitution of these $N$ solutions into the first equation yields the second-order wave equation,
\begin{equation}
\Biggl[ \left( \hat{E} - U \right) I - \frac{1}{2m_1} \left( \hat{P}_1 - \frac{q_1}{c} A_1 \right)^2 - 
\cdots - \frac{1}{2m_N} \left( \hat{P}_N - \frac{q_N}{c} A_N \right)^2 \Biggr] {\underline\psi}_1 = 0 ,
\label{eq:3.29}
\end{equation} 
for the $N$ non-relativistic particles of the spinor wave function ${\underline\psi}_1(\underline{x}_1, \underline{x}_2, \cdots, \underline{x}_N, t)$.

Next the energies are found for the interaction of the magnetic fields with the intrinsic magnetic spin moments of the $N$ particles.  The techniques used to accomplish this are adaptations of the Levy-Leblond approach first developed in [4] and in more detail in [10].

To find the interaction terms of the intrinsic magnetic moments of the $N$ particles with the magnetic fields experienced by the particles one needs to extend the following well-known identity.  For a single spin-1/2 particle,
\begin{equation}
AB = {\underline\alpha} \cdot {\underline\beta} + i \hat{\sigma} \cdot ({\underline\alpha} \times {\underline\beta})
\label{eq:3.30}
\end{equation}
where $\hat\sigma$ is the Pauli spin-vector operator in (\ref{eq:2.5}), ${\underline\alpha} = [\alpha_1, \alpha_2, \alpha_3]$ and ${\underline\beta} = [\beta_1, \beta_2, \beta_3]$ are two 3-vectors of operators, and
\begin{eqnarray}
A &\equiv& \hat{\sigma} \cdot {\underline\alpha} = \left[ \begin{array}{cc} \alpha_3 & \alpha_1 - i \alpha_2 \\
  \alpha_1 + i \alpha_2 & - \alpha_3 \end{array} \right] , \nonumber\\
B &\equiv& \hat{\sigma} \cdot {\underline\beta} = \left[ \begin{array}{cc} \beta_3 & \beta_1 - i \beta_2 \\
  \beta_1 + i \beta_2 & - \beta_3 \end{array} \right] .
\label{eq:3.31}
\end{eqnarray}
The identity in (\ref{eq:3.30}) is easily seen to apply in the $N$-particle case as well, by the definition (\ref{tensor_product_spin}) of the $k$th spin operator $\hat\sigma$, where now we define $A_k \equiv \hat\sigma_k \cdot {\underline\alpha}_k$ and $B_k  \equiv \hat\sigma_k \cdot {\underline\beta}_k$, and the identity becomes $A_k B_k = {\underline\alpha}_k \cdot {\underline\beta}_k + i \hat{\sigma}_k \cdot ({\underline\alpha}_k \times {\underline\beta}_k)$. If we apply it to the $k$th squared matrix of the last $N$ terms of operators in the partial differential equation in (\ref{eq:3.29}), we get
\begin{equation}
\begin{split}
\left( \hat{P}_k - \frac{q_k}{c} A_k \right)^2 =
  \left( \hat{\underline{p}}_k - \frac{q_k}{c} {\underline{A}}_k \right) \cdot
  \left( \hat{\underline{p}}_k - \frac{q_k}{c} {\underline{A}}_k \right)  \\
+ i \hat\sigma_k \cdot \left[ \left( \hat{\underline{p}}_k - \frac{q_k}{c} {\underline{A}}_k \right) \times
  \left( \hat{\underline{p}}_k - \frac{q_k}{c} {\underline{A}}_k \right) \right] .
\end{split}
\label{eq:3.32}
\end{equation}
By (\ref{eq:3.25}) $\hat{\underline{p}}_k$ is a differential operator which does not necessarily commute with a function.  Thus the term in the brackets when applied to the wave function  $\hat{\underline\psi}_1$, is simplified as follows:
\begin{equation}
\begin{split}
\left( \hat{\underline{p}}_k - \frac{q_k}{c} {\underline{A}}_k \right) \times&
  \left( \hat{\underline{p}}_k - \frac{q_k}{c} {\underline{A}}_k \right) {\underline\psi}_1 \\
= \Biggl[ \hat{\underline{p}}_k \times \hat{\underline{p}}_k - \frac{q_k}{c}({\underline{A}}_k \times \hat{\underline{p}}_k )
&- \frac{q_k}{c}( \hat{\underline{p}}_k \times {\underline{A}}_k )
+ \frac{q^2_k}{c^2}( {\underline{A}}_k \times {\underline{A}}_k ) \Biggr] {\underline\psi}_1 \\
= - \frac{q_k}{c}({\underline{A}}_k \times \hat{\underline{p}}_k ) {\underline\psi}_1 - 
& \frac{q_k}{c}\left[ ( \hat{\underline{p}}_k \times {\underline{A}}_k ) - ({\underline{A}}_k \times \hat{\underline{p}}_k ) \right] {\underline\psi}_1  \\
=& - \frac{q_k}{c}( \hat{\underline{p}}_k \times {\underline{A}}_k ) {\underline\psi}_1 . \nonumber
\end{split}
\end{equation}
Hence a substitution of this result into (\ref{eq:3.32}) yields the identity,
\begin{eqnarray}
\left( \hat{P}_k - \frac{q_k}{c} A_k \right)^2
&=& \left( \hat{\underline{p}}_k - \frac{q_k}{c} {\underline{A}}_k \right)^2
 - i\frac{q_k}{c}\left[\hat\sigma_k \cdot ( \hat{\underline{p}}_k \times {\underline{A}}_k ) \right] \nonumber\\
&=& \left( \hat{\underline{p}}_k - \frac{q_k}{c} {\underline{A}}_k \right)^2
 - \frac{\hbar q_k}{c} \hat\sigma_k \cdot {\underline{B}}_k
\label{eq:3.33}
\end{eqnarray} 
for $k=1,2,\cdots,N$ since $\nabla_k \times {\underline{A}}_k = {\underline{B}}_k$ is the magnetic field in the vicinity of the $k$-th particle.  This result is the same as that obtained in [10, Sec. 13.2] for $N=1$.

Finally a use of (\ref{eq:3.33}) in (\ref{eq:3.29}) and a replacement of ${\underline\psi}_1$ by $\underline\psi$ yields the following general Schr\"odinger-like equation,
\begin{equation}
\Biggl( \left[ \hat{E} - U - \sum_{k=1}^N \frac{1}{2m_k}\left( \hat{\underline{p}}_k - \frac{q_k}{c}{\underline{A}}_k \right)^2 \right]
  + \sum_{k=1}^N \frac{q_k\hbar}{2m_k c} (\hat\sigma_k \cdot {\underline{B}}_k ) \Biggr)
  {\underline\psi} = 0 ,
\label{eq:3.34}
\end{equation} 
or its equivalent in differential-operator form,
\begin{equation}
\Biggl( \left[ i\hbar\frac{\partial}{\partial t} - U - \sum_{k=1}^N \frac{1}{2m_k}\left( \frac{\hbar}{i}\nabla_k - \frac{q_k}{c}{\underline{A}}_k \right)^2 \right]
+ \sum_{k=1}^N \frac{q_k\hbar}{2m_k c} (\hat\sigma_k \cdot {\underline{B}}_k ) \Biggr)
  {\underline\psi} = 0 .
\label{eq:3.35}
\end{equation} 
For $N=1$ the wave equation in (\ref{eq:3.35}) agrees exactly with the Pauli equation with spin, e.g., see [10, Secs. 12.5 and 13.2], where
\[
\frac{q\hbar}{2mc} \hat\sigma \cdot {\underline{B}} = \hat{\underline\mu} \cdot {\underline{B}}
\]
is the magnetic interaction energy of ${\underline{B}}$ with the intrinsic magnetic moment of the particle being
\begin{equation}
\hat{\underline\mu}= \frac{q\hbar}{2mc} \hat\sigma = \frac{q\hbar}{mc} \hat{\underline{S}}
\label{eq:3.36}
\end{equation}
in terms of the usual spin operator, $\hat{\underline{S}} = (\frac{1}{2}) \hat\sigma$.  Thus for $N$ particles the total spin magnetic-interaction energy operator is given by
\begin{equation}
\hat{E}_{\rm spin} = \sum_{k=1}^N \frac{q_k\hbar}{2m_k c} (\hat\sigma_k \cdot {\underline{B}}_k )
  = \sum_{k=1}^N \hat{\underline\mu}_k \cdot {\underline{B}}_k
\label{eq:3.37}
\end{equation}
in terms of the intrinsic magnetic moments ${\underline{B}}_k$, defined in (\ref{eq:3.36}).  Finally if ${\underline{B}}$ is a uniform magnetic field, eq.~(\ref{eq:3.35}) reduces to the Schr\"odinger equation in [9, Sec. 113] for an atom of $N$ electrons.  (Note that here we are neglecting dipole interactions {\it between} the different spins.)

In the next section the general Pauli-Schr\"odinger wave equation for $N$ particles is applied to the simple Zeeman effect for the hydrogen atom in a weak magnetic field.  The problem is treated as a two-body problem, an electron and a proton, both with spin $\frac{1}{2}$, rather than as a one-body problem of a single electron with a nucleus of a large mass, e.g., see [10, Ex. 12.3].

\section{\nothuge The Weak Magnetic-Field Zeeman Effect for Two Oppositely Charged Particles}

In this section the splitting of the spectral lines in a weak magnetic field of a hydrogen-like system is studied using the generalized Paul equation in (\ref{eq:3.34}) for $N=2$ particles.  This new treatment yields the simple Zeeman effect with a slightly modified Larmor frequency which depends on the difference of the masses of the two particles.  The simple Zeeman effect given here is similar to the classical treatment, which uses the Pauli equation for $N=1$, in that the relativistic spin-orbit interactions are neglected, e.g., see [10, Ex. 12.3].

In the Zeeman effect the magnetic field $\underline{B}$ is assumed to be homogeneous.  Hence there is no loss in generality to let only the $z$ component of $\underline{B}$ be nonzero, i.e.,
\begin{equation}
\underline{B} = (0,0,B),
\label{eq:4.1}
\end{equation}
where $B$ is a constant.  Since the vector potential $\underline{A}$ is related to $\underline{B}$ by $\underline{B} = \nabla\times\underline{A}$, the quantity,
\begin{equation}
\underline{A} = \left( - \frac{1}{2} By, \frac{1}{2} Bx, 0 \right) ,
\label{eq:4.2}
\end{equation}
is chosen to be the vector potential associated with $\underline{B}$ in (\ref{eq:4.1}).

The generalized Pauli equation in (\ref{eq:3.34}) or (\ref{eq:3.35}) for $N=2$ has the form,
\begin{eqnarray}
\hat{H} \underline\psi = \Biggl[ \frac{1}{2m_1} \left( \hat{\underline{p}}_1 - \frac{q_1}{c}\underline{A}_1 \right)^2
+  \frac{1}{2m_2} \left( \hat{\underline{p}}_2 - \frac{q_2}{c}\underline{A}_2 \right)^2  \\
- \sum_{k=1}^2 \frac{q_k\hbar}{2m_k c} (\hat\sigma_k \cdot \underline{B}_k) + U \Biggr] \underline\psi
  = i \hbar \frac{\partial}{\partial t}\underline\psi ,\nonumber
\label{eq:4.3}
\end{eqnarray}
where
\begin{equation}
\underline{A}_k = \left( - \frac{1}{2} By_k, \frac{1}{2} Bx_k, 0 \right) ,
\label{eq:4.4}
\end{equation}
for $(k=1,2)$ and 
\begin{equation}
\underline{B}_1 = \underline{B}_2 = \underline{B}
\label{eq:4.5}
\end{equation}
with $\underline{B}$ being the uniform magnetic field given in (\ref{eq:4.1}).  Eq.~(\ref{eq:4.3}) is used in the ensuing discussion to study the simple Zeeman effect of a hydrogen-like atom of two particles, the negatively charged electron and the positively charged nucleus.  Here in (\ref{eq:4.3}) one assumes that the charges and masses of these two particles are, respectively,
\begin{equation}
q_1 = -e, \ \ q_2 = +e
\label{eq:4.6}
\end{equation}
and $m_1$, $m_2$, where $e=|e|$ is the magnitude of the charge of an electron and $m_1 \le m_2$.  Finally the potential $U$ in (\ref{eq:4.3}) is the Coulomb potential given by
\begin{equation}
U = - \frac{Ze^2}{r} ,
\label{eq:4.7}
\end{equation}
where
\begin{equation}
r = |\underline{r}| = |\underline{r}_1-\underline{r}_2|
\label{eq:4.8}
\end{equation}
is the magnitude of the relative position vector $\underline{r}$, given by
\begin{equation}
\underline{r} = \underline{r}_1 - \underline{r}_2
\label{eq:4.9}
\end{equation}
with $\underline{r}_k$ for $k=1,2$ being the position vectors of the $k$-th particle.

The spin degrees of freedom for this system have a 4-dimensional Hilbert space.  This can be expressed separately in terms of the spins of the two particles; but one can switch to a different representation in which the magnetic moments are parallel or antiparallel.    We will simplify the equations by assuming that the magnetic moments of the two spins are oppositely directed, i.e., we restrict ourselves to single two-dimensional subspace.  In this case, we can treat the system as if it had a single spin-1/2 degree of freedom.  We will denote the effective operator for this spin by $\hat\sigma$.  (This is energetically favored by the spin-spin interaction, though we are otherwise neglecting that effect in this section.)

Since the magnetic field $\underline{B}$ in (\ref{eq:4.1}) is assumed to be weak, terms of order $\underline{A}_k^2 = \underline{A}_k\cdot\underline{A}_k$ in (\ref{eq:4.3}) are neglected, and one obtains the approximate wave equation,
\begin{eqnarray}
\hat{H} \underline\psi &=& \biggl\{ \frac{1}{2M} \biggl[ \frac{1}{\mu_1} \biggl( \underline{\hat{p}}_1^2 + \frac{2e}{c} \underline{A}_1 \cdot \underline{\hat{p}}_1\biggr)
 + \frac{1}{\mu_2} \biggl( \underline{\hat{p}}_2^2 + \frac{2e}{c} \underline{A}_2 \cdot \underline{\hat{p}}_2\biggr)  \nonumber\\ 
&& + \frac{e\hbar}{c}\biggl( \frac{1}{\mu_1} - \frac{1}{\mu_2} \biggr)(\hat\sigma\cdot\underline{B})\biggr] + \frac{Ze^2}{r} \biggr\}\underline\psi
= i\hbar \frac{\partial}{\partial t} \underline{\psi} ,
\label{eq:4.10} 
\end{eqnarray}
where
\begin{equation}
\mu_k = m_k/M
\label{eq:4.11}
\end{equation}
are the normalized particle masses for $k=1,2$, and
\begin{equation}
M = m_1 + m_2
\label{eq:4.12}
\end{equation}
is the total mass of the two particles.

The position vectors $\underline{r}_k$ of the two particles are transformed next into the relative position vector $\underline{r}$ in (\ref{eq:4.9}) and the center-of-mass position vector $\underline{R}$ by the following two relations:
\begin{equation}
\underline{r} = \underline{r}_1 - \underline{r}_2 \ \ \ {\rm and} \ \ \ 
\underline{R} = \mu_1 \underline{r}_1 + \mu_2 \underline{r}_2 ,
\label{eq:4.13}
\end{equation}
where $\mu_1$ and $\mu_2$ are the normalized masses of the two particles, defined in (\ref{eq:4.11}).  This transformation of coordinates, using (\ref{eq:3.25}) and the rules of partial differentiation, yields readily
\begin{equation}
\underline{\hat{p}}_1 = \underline{\hat{p}} + \mu_1\underline{\hat{P}} \ \ \ {\rm and} \ \ \  
\underline{\hat{p}}_2 = - \underline{\hat{p}} + \mu_2\underline{\hat{P}}
\label{eq:4.14}
\end{equation}
as the transformation of the momentum operators $\underline{\hat{p}}_1$ and $\underline{\hat{p}}_2$ onto the momentum operators $\underline{\hat{p}}$ and $\underline{\hat{P}}$.  Here specifically
\begin{equation}
\underline{\hat{p}} = \frac{\hbar}{i} \mu \nabla_{\underline{r}}
\label{eq:4.15}
\end{equation}
is the relative-position momentum operator in terms of the classical reduced mass $\mu$, where
\begin{equation}
\frac{1}{\mu} = \frac{1}{m_1} + \frac{1}{m_2} = \frac{1}{M} \left( \frac{1}{\mu_1} + \frac{1}{\mu_2} \right) ,
\label{eq:4.16}
\end{equation}
and $\nabla_{\underline{r}}$ is the gradient operator with respect to the components of $\underline{r}$.  Also
\begin{equation}
\underline{\hat{P}} = \frac{\hbar}{i} M \nabla_{\underline{R}}
\label{eq:4.17}
\end{equation}
is the center-of-mass momentum operator in terms of the total mass $M$ in (\ref{eq:4.12}) and the gradient operator $\nabla_{\underline{R}}$ with respect to the components of $\underline{R}$.  Also a substitution of the relatoins in (\ref{eq:4.4}) produces the vector potentials,
\begin{eqnarray}
\underline{A}_1 &=& \underline{A}_{\underline{R}} + \mu_2 \underline{A}_{\underline{r}} \nonumber\\
\underline{A}_2 &=& \underline{A}_{\underline{R}} - \mu_1 \underline{A}_{\underline{r}} ,
\label{eq:4.18}
\end{eqnarray}
in terms of the new vector potentials,
\begin{eqnarray}
\underline{A}_{\underline{r}} &=& \frac{B}{2}(\mu_2 - \mu_1)[-y,x,0] \nonumber\\
\underline{A}_{\underline{R}} &=& \frac{B}{2}(\mu_2 - \mu_1)[-Y,X,0].
\label{eq:4.19}
\end{eqnarray}
Evidently $\underline{A}_{\underline{r}}$ and $\underline{A}_{\underline{R}}$ are functions of the relative and center-of-mass coordinates, respectively.  Finally a substitution of the momentum operator relations in (\ref{eq:4.14}) and the vector potentials in (\ref{eq:4.19}) yields the Pauli-like wave equation
\[
\frac{1}{2M}\biggl[ \frac{1}{\mu_1\mu_2} \left( \underline{\hat{p}}^2 + \frac{2e}{c}\left[ (\mu_2 - \mu_1)\underline{A}_{\underline{r}} + \underline{A}_{\underline{R}} \right] \cdot \underline{\hat{p}} \right)
+ \left((\underline{\hat{P}}^2 + \frac{2e}{c})\underline{A}_{\underline{r}}\cdot \underline{\hat{P}} \right)
\]
\begin{equation}
+ \frac{e\hbar B}{c} \left( \frac{1}{\mu_1} - \frac{1}{\mu_2} \right) \sigma - \frac{e^2}{r} \biggr] \underline\psi 
= i\hbar \frac{\partial}{\partial t} \underline\psi ,
\label{eq:4.20}
\end{equation}

for two particles in a weak magnetic field in the relative and center-of-mass coordinates.

Now express equation (\ref{eq:4.20}) in spinor form as follows:
\begin{equation}
\hat{H} \underline\psi = \left[ \begin{array}{cc} \hat{H}_1 & 0 \\ 0 & \hat{H}_2 \end{array} \right]
\left[ \begin{array}{c} \psi_1 \\ \psi_2 \end{array} \right] = i\hbar \frac{\partial}{\partial t} \left[ \begin{array}{c} \psi_1 \\ \psi_2 \end{array} \right] ,
\label{eq:4.21}
\end{equation}
where
\begin{subequations}
\begin{eqnarray}
\hat{H}_1 &=& \hat{H}'_{\underline{r}} + \hat{H}'_{\underline{R}}
  + \frac{e\hbar B}{2c} \left[\frac{1}{m_1} - \frac{1}{m_2} \right] ,\nonumber\\ \label{eq:4.22a}\\
\hat{H}_2 &=& \hat{H}'_{\underline{r}} + \hat{H}'_{\underline{R}}
  - \frac{e\hbar B}{2c} \left[\frac{1}{m_1} - \frac{1}{m_2} \right] .\nonumber\\ \label{eq:4.22b}
\end{eqnarray}
\end{subequations}
Here also from (\ref{eq:4.20}) one has
\begin{subequations}
\begin{eqnarray}
\hat{H}'_{\underline{r}} &=& \frac{1}{2\mu} \left[ \underline{\hat{p}} + \frac{e}{c} (\mu_2 - \mu_1) \underline{A}'_{\underline{r}} \right]^2 - \frac{e^2}{r} \nonumber\\
&\approx& \frac{1}{2\mu} \left[ \underline{\hat{p}}^2 + \frac{e}{c} (\mu_2 - \mu_1) \left( \underline{A}'_{\underline{r}} \cdot \underline{\hat{p}} \right) \right] - \frac{e^2}{r}\nonumber\\
\label{eq:4.23a}
\end{eqnarray}
and
\begin{eqnarray}
\hat{H}'_{\underline{r}} &=& \frac{1}{2M} \left[ \underline{\hat{P}} + \frac{e}{c} \underline{A}'_{\underline{R}} \right]^2 \nonumber\\
&\approx& \frac{1}{2M} \left[ \underline{\hat{P}}^2 + \frac{e}{c} \left( \underline{A}'_{\underline{R}} \cdot \underline{\hat{P}} \right) \right] ,\nonumber\\
\label{eq:4.23b}
\end{eqnarray}
\end{subequations}
where
\begin{subequations}
\begin{eqnarray}
\underline{A}'_{\underline{r}} &=& \underline{A}_{\underline{r}} + \frac{1}{\mu_2-\mu_1} \underline{A}_{\underline{R}} \nonumber\\
&=& \underline{A}_{\underline{r}} + \nabla_{\underline{r}} \sigma(\underline{r},\underline{R}) , \label{eq:4.24a} \\
\underline{A}'_{\underline{R}} &=& 0 + \underline{A}_{\underline{r}} = \nabla_{\underline{R}} \tau(\underline{R},\underline{r}) , \label{eq:4.24b}
\end{eqnarray}
\end{subequations}
are the gauge transformations of the vector potentials, $\underline{A}_{\underline{r}}$ and $\underline{A}_{\underline{R}} = 0$ associated with the apparent interactions energies of the linear and angular momenta of the system with
\begin{subequations}
\begin{eqnarray}
\sigma(\underline{r}; \underline{R}) &=& \frac{1}{\mu_2-\mu_1} ( \underline{r} \cdot \underline{A}_{\underline{R}} ) , \label{eq:4.25a} \\
\tau(\underline{R}; \underline{r}) &=& \underline{R} \cdot \underline{A}_{\underline{r}} \label{eq:4.25b}
\end{eqnarray}
\end{subequations}
being, using (\ref{eq:4.20}), the corresponding gauge functions.

In order to separate the variables in the vector partial differential equations in (\ref{eq:4.21}) let
\begin{subequations}
\begin{equation}
\psi_k = f_k(t) g'_k(\underline{R}; \underline{r}) \phi'_k( \underline{r};  \underline{R})
\label{eq:4.26a}
\end{equation}
be the factorization of $\psi_k$, where
\begin{eqnarray}
g'_k &=& g_k(\underline{R}) \exp \left[ \frac{-ie}{\hbar c} \tau(\underline{R};\underline{r}) \right] ,\nonumber\\
\phi'_k &=& \phi_k(\underline{R}) \exp \left[ \frac{-ie}{\hbar c} \sigma(\underline{R};\underline{r}) \right] 
\label{eq:4.26b}
\end{eqnarray}
\end{subequations}
for $k=1,2$ with $f_k(t)$, $g_k(\underline{R})$, $\psi_k(\underline{r})$ being, respectively, functions only of $t$, $\underline{R}$, and $\underline{r}$. Also the gauge functions $\tau$ and $\sigma$ are defined in (\ref{eq:4.25a}) and (\ref{eq:4.25b}).  A substitution of (\ref{eq:4.26a}) into (\ref{eq:4.21}) yields
\[
\hat{H}_k f_k g'_k \phi'_k = i\hbar \frac{\partial}{\partial t} f_k g'_k \phi'_k = (i\hbar{\dot f}_k) g'_k \phi'_k
\]
so that one obtains the ratios,
\begin{equation}
\frac{\hat{H}_k g'_k \phi'_k}{g'_k \phi'_k} = \frac{i\hbar\dot{f}_k}{f_k} = E_T ,
\label{eq:4.27}
\end{equation}
where $E_T$ is constant for all $t, \underline{R}, \underline{r}$ and $k=1,2$.  Here $E_T$ is identified with the total energy of the two-particle system in the magnetic field $\underline{B}$.  From the second equality in (\ref{eq:4.27}) one obtains
\[
f_k(t) = e^{-\frac{i}{\hbar} E_T t}
\]
for the time function $f_k(t)$.

A use of (\ref{eq:4.23a}) and (\ref{eq:4.23b}) in the terms of equation (\ref{eq:4.27}), when rearranged, yields the equations,
\[
\left( \hat{H}_1 - E_T \right) g'_1 \phi'_1 = \Biggl( \hat{H}'_{\underline{r}} + \hat{H}'_{\underline{R}}
+ \frac{e\hbar B}{2c} \left[ \frac{1}{m_1} - \frac{1}{m_2} \right] - E_T \Biggr) g'_1 \phi'_1 ,
\]
\[
\left( \hat{H}_2 - E_T \right) g'_2 \phi'_2 = \Biggl( \hat{H}'_{\underline{r}} + \hat{H}'_{\underline{R}}
- \frac{e\hbar B}{2c} \left[ \frac{1}{m_1} - \frac{1}{m_2} \right] - E_T \Biggr) g'_2 \phi'_2 . \nonumber
\]
From this pair of equations one obtains the ratios,
\begin{subequations}
\begin{eqnarray}
\frac{ - \left[ \hat{H}'_{\underline{r}} + \frac{e\hbar B}{2c} \left( \frac{1}{m_1} - \frac{1}{m_2} \right) - E_T \right] \phi'_1 }{\phi'_1} &=& \frac{\hat{H}'_{\underline{R}} g'_1 }{g'_1} = E_L , \label{eq:4.28a} \\
\frac{ - \left[ \hat{H}'_{\underline{r}} - \frac{e\hbar B}{2c} \left( \frac{1}{m_1} - \frac{1}{m_2} \right) - E_T \right] \phi'_2 }{\phi'_2} &=& \frac{\hat{H}'_{\underline{R}} g'_2 }{g'_2}  = E_L ,  \label{eq:4.28b}
\end{eqnarray}
\end{subequations}
where $E_L$ is again a constant, but in this case for all $\underline{R}$, $\underline{r}$ and $k=1,2$.  $E_L$ is associated with the translational energy of the system.

The second equalities in (\ref{eq:4.28a}) and (\ref{eq:4.28b}) yield immediately by (\ref{eq:4.23a}) the result,
\begin{equation}
\frac{1}{2M} \left[ \underline{\hat{P}} + \frac{e}{c} \underline{A}'_{\underline{R}} \right]^2 g'_k
+ E_L g'_k = 0,
\label{eq:4.29}
\end{equation}
for the motion of the center-of-mass of the two-particle system.  By (\ref{eq:4.26b}) and the gauge transformation in (\ref{eq:4.24b}) one finds the identity,
\begin{eqnarray}
\left[ \underline{\hat{P}} + \frac{e}{c} \underline{A}'_{\underline{R}} \right] g'_k
&=& \left[ \frac{\hbar}{i} \nabla_{\underline{R}} + \frac{e}{c} \nabla_{\underline{R}} \tau \right] g'_k
\exp\left( \frac{-ie}{\hbar c} \tau \right)
\nonumber\\
&=& \left[\exp\left( \frac{-ie}{\hbar c} \tau \right) \right] \underline{\hat{P}} g'_k , \nonumber
\end{eqnarray}
which transforms (\ref{eq:4.29}) into the Schr\"odinger-equation pair,
\[
\frac{\hbar^2}{2M} \nabla^2_{\underline{R}} g_k + E_L g_k = 0, \ \ \ (k=1,2) ,
\]
for the translational motion of the two-particle system.  See [10, Sec. 9.1] for an excellent discussion of the gauge invariance of Schr\"odinger's equation, except for the usually unwanted phase factor.

In a similar fashion the first equalities in (\ref{eq:4.28a}) and (\ref{eq:4.28b}) produce the wave equation pair,
\begin{subequations}
\begin{equation}
\left[ \hat{H}_{\underline{r}} + \frac{eB}{2c} \left( \frac{1}{m_1} - \frac{1}{m_2} \right) \sigma_3 \right] \phi_k = E \phi_k ,
\label{eq:4.30a}
\end{equation}
for $k=1,2$, where
\begin{eqnarray}
\hat{H}_{\underline{r}} &=& \frac{1}{2\mu} \left[ \underline{\hat{p}} + \frac{e}{c} (\mu_2 - \mu_1) \underline{A}_{\underline{r}} \right]^2 - \frac{e^2}{r} \nonumber\\
&\approx& \frac{1}{2\mu} \left[ \underline{\hat{p}}^2 + \frac{e}{c} ( \mu_2 - \mu_1 ) ( \underline{A}_{\underline{r}} \cdot \underline{\hat{p}} ) \right] - \frac{e^2}{r} ,\nonumber\\
\label{eq:4.30b}
\end{eqnarray}
\end{subequations}
is the gauge-transformed wave equation in the relative coordinates $\underline{r}$ of the two-particle system.  Here also $E = E_T - E_L$ is the total rotational energy of the system.

Next observe by (\ref{eq:4.2}) that the term $\underline{A}_{\underline{r}} \cdot \underline{\hat{p}}$ in (\ref{eq:4.30a}) is given by
\begin{equation}
\underline{A}_{\underline{r}} \cdot \underline{\hat{p}} = i\hbar \frac{B}{2} \left( y \frac{\partial}{\partial x} - x \frac{\partial}{\partial y} \right) = \frac{B}{2} \hat{L}_z ,
\label{eq:4.31}
\end{equation}
where $\hat{L}_z$ is the $z$-component of the angular-momentum operator, e.g., see [8, Sec 4.8].  Thus a substitution of (\ref{eq:4.31}) into (\ref{eq:4.30b}) and thence into (\ref{eq:4.30a}) yields the wave equation,
\begin{equation}
\hat{H}_0 \underline\phi + \frac{eB}{2c} \left( \frac{1}{m_1} - \frac{1}{m_2} \right) ( \hat{L}_z + \hbar\sigma_3) \underline\phi = E \underline\phi ,
\label{eq:4.32}
\end{equation}
where
\begin{equation}
\hat{H}_0 = \underline{\hat{p}}^2/2\mu + e^2/r
\label{eq:4.33}
\end{equation}
and $e$ is the magnitude of the charge of an electron.

Now define a two-particle Larmor frequency as follows:
\begin{equation}
\omega_L = \frac{eB}{2c} \left( \frac{1}{m_1} - \frac{1}{m_2} \right) = \frac{eB}{2cm_L} ,
\label{eq:4.34}
\end{equation}
where $m_L$ is the ``increased'' mass, given by
\begin{equation}
\frac{1}{m_L} = \frac{1}{m_1} - \frac{1}{m_2} ,
\label{eq:4.35}
\end{equation}
needed in the orbital magnetic-moment energy operator, $eB\hat{L}_z/2cm_L$.  In terms of the new Larmor frequency in (\ref{eq:4.34}) the wave equation in (\ref{eq:4.32}) is expressed finally by
\begin{equation}
\hat{H}_0 \underline\phi + \omega_L ( \hat{L}_z + \hbar\sigma_3) \underline\phi = E \underline\phi ,
\label{eq:4.36}
\end{equation}
or in spin components by the pair,
\begin{subequations}
\begin{eqnarray}
\hat{H}_0 \phi + \omega_L ( \hat{L}_z + \hbar\sigma_3) \phi_1 &=& E \phi_1 , \label{eq:4.37a} \\
\hat{H}_0 \phi + \omega_L ( \hat{L}_z + \hbar\sigma_3) \phi_2 &=& E \phi_2 . \label{eq:4.37b}
\end{eqnarray}
\end{subequations}

If the magnetic induction $B$ is zero, i.e., $B=0$, then
\begin{equation}
\phi_1 = \phi_2 = \psi_{nlm} = R_{nl} Y_{lm} (\theta, \phi)
\label{eq:4.38}
\end{equation}
are the well-known standard eigensolutions of the pair of equations (\ref{eq:4.37a}) and (\ref{eq:4.37b}), e.g., see [10, Sec. 9.2], where here by (\ref{eq:4.16}) mass $m$ is replaced by the reduced mass $\mu$.  Note also that the wave function $\psi_{nlm}$ in (\ref{eq:4.38}) is an eigenfunction of $\hat{L}_z$, i.e.,
\begin{equation}
\hat{L}_z \psi_{nlm} = \hbar m \psi_{nlm} ,
\label{eq:4.39}
\end{equation}
e.g., see [10, pp. 67 and 245].

Now let $E_{nlm}$ be the eigenvalues of the eigenfunction $\psi_{nlm}$ of $\hat{H}_0$ in (\ref{eq:4.33}) so that one has
\begin{equation}
\hat{H}_0 \psi_{nlm} = E_{nlm} \psi_{nlm} .
\label{eq:4.40}
\end{equation}
Thus putting
\begin{equation}
\underline\psi^{(1)}_{nlm} = \left[ \begin{array}{c} \psi_{nlm} \\ 0 \end{array} \right]
\label{eq:4.41}
\end{equation}
into the left side of (\ref{eq:4.36}) yields by (\ref{eq:4.39}) and (\ref{eq:4.40})
\[
\hat{H}_0 \left[ \begin{array}{c} \psi_{nlm} \\ 0 \end{array} \right]
+ \omega_L( \hat{L}_z + \hbar) \left[ \begin{array}{c} \psi_{nlm} \\ 0 \end{array} \right]
 = [E_{nlm} + \omega_L \hbar (m+1) ] \left[ \begin{array}{c} \psi_{nlm} \\ 0 \end{array} \right] .
\] 
Hence $\underline\psi^{(1)}_{nlm}$ in (\ref{eq:4.41}) is a spinor eigenfunction of the two-particle wave equation in (\ref{eq:4.36}) with eigenvalue,
\begin{equation}
E^{(1)}_{nlm} = E_{nlm} + \omega_L \hbar (m+1) .
\label{eq:4.42}
\end{equation}
Similarly
\begin{equation}
E^{(1)}_{nlm} = E_{nlm} + \omega_L \hbar (m-1)
\label{eq:4.43}
\end{equation}
is the eigenvalue of the spinor eigenfunction,
\[
\underline\psi^{(2)}_{nlm} = \left[ \begin{array}{c} 0 \\ \psi_{nlm} \end{array} \right] .
\]
The figure in [10, pg. 248] illustrates this two-fold splitting of the energy levels of a hydrogen-like atom for $n=1,2$.

The above application of the $N$-particle Pauli equation in (\ref{eq:3.37}) to the Zeeman effect for $N=2$ evidently reduces tot he corresponding problem for $N=1$ with a reduced mass $\mu$, except for the modified Larmor frequency $\omega_L$, given in (\ref{eq:4.34}).  The fact that the Larmor frequency $\omega_L$ in (\ref{eq:4.34}) is expressed in terms of an increased mass $m_L$ in (\ref{eq:4.35}), instead of the usual electron mass, appears to have been noted previously for two special cases.

In 1952 W. H. Lamb in [11], investigated by a use of the Dirac equation for a relativistic particle in a field the effect of nuclear motion ont he orbital energy of theZeeman effect of a hydrogen-like atom.  He found in his eq. (153) that the effective $g$-value for orbital motion, taking the motion of the nucleus into account, is given by
\[
g_L = m_1^{-1} - m_2^{-1} .
\]
Lamb's effective $g$-value is equivalent to the inverse in (\ref{eq:4.35}) of $m_L$, the ``increased'' mass.

Finally it is known that in a weak magnetic field there is no Zeeman effect in positronium, e.g., see [12, pg. 214].  This agrees with the result obtained by letting $m_1 = m_2 = m$ be the common mass of both the electron and  positron, so that by (\ref{eq:4.16}) and (\ref{eq:4.34}) $\mu = m/2$ and $\omega_L = 0$.  Thus, ignoring higher-order relativistic corrections, positronium has no weak-field Zeeman effect.

The above observations agree with the new, more complete, Larmor frequency obtained in (\ref{eq:4.34}).  This in turn suggests the validity of the generalized Pauli equation in (\ref{eq:3.34}) and the power of the techniques developed in Sections 1 and 2 to obtain this equation.

\section{\nothuge Concluding Remarks}

The techniques developed herein for the general Pauli wave equation with spin generalize the results obtained by Levy-Le Blond in [4], but also the results in [8].  In [8], a similar, but special $4\times4$ matrix is singular if and only if a certain relativistic Hamiltonian relation is zero for a free relativistic particle.  A procedure, similar to that developed in this paper in Section 1, is used in [8] to demonstrate the existence of vectro solutions $\underline\theta(\underline{p},\underline{q})$ of $G\underline\theta = 0$ in the null space $N_G$ of $G$, where $\underline{p} = [\frac{E}{c}, p_1, p_2, p_3]$ is the energy-momentum vector of special relativity.  However, in [8] the properties of $N_G$ are found in considerably more detail by an actual construction, using the special structure of the $4\times4$ matrix $G$.

The quantization procedure, generalized in Section 2, is applied in [8] to the equation, $G\underline\theta = 0$, to obtain Dirac's equation for a free particle.  The Fourier-momentum transforms of the functions in the null space $N_G$ are shown in [8] to be the general solutions of Dirac's wave equation.  It is believed that the more general methods developed here in Sections 1 and 2, using the extended Hamiltonian, can be generalized to other problems of relativistic quantum physics.

Related to this is the possibility that a higher-order Hermitian matrix $G$ can be found which incorporates a finite number of relativistic correction terms of the series for $H$ in (\ref{eq:3.1}).  With such a larger matrix, it might be possible to use the procedures of Sections 1 and 2 to find relativistically corrected wave equations for $N$ particles with spin.  However, the extra complexity, needed to find these equations, requires that this relativistic extension to be accomplished elsewhere.

\section*{\nothuge Appendix A}

The formula in (\ref{eq:3.13}) to evaluate the determinant of the $n\times n$ matrix $M$ in (\ref{eq:3.12}) in block form results from the following factorization lemma for partitioned matrices.

{\bf Lemma 1}   Let $A$ and $D$ be, respectively, $k\times k$ and $(n-k)\times(n-k)$ square block matrices with $D$ being non-singular.  Then the identity,
\[
M = \left[ \begin{array}{cc} A & B \\ C & D \end{array} \right]
\]
\[
= \left[ \begin{array}{cc} I_k & BD^{-1} \\ 0 & I_{n-k} \end{array} \right] 
 \left[ \begin{array}{cc} A - B D^{-1} C & 0 \\ 0 & D \end{array} \right] 
 \left[ \begin{array}{cc} I_k & 0 \\ D^{-1} C & I_{n-k} \end{array} \right] ,\nonumber
\] 
holds, where $B$ and $D$ are, respectively, $k\times(n-k)$ and $(n-k)\times k$ rectangular block matrices.

{\bf Proof:}  This lemma is verified by a direct multiplication of the three matrix factors in the identity.  The determinantal identity in (ref{eq:3.13}) is obtained immediately by using the fact that the determinant of a product of matrices is the product of their determinants and Laplace's rules for evaluating determinants.

\end{document}